# Coherence properties of photons emitted by single defect centers in diamond

A. Batalov[1], C. Zierl[1], T. Gaebel[1], P. Neumann[1], I.-Y. Chan[1], G. Balasubramanian[1], P. R. Hemmer[2], F. Jelezko[1]*, and J. Wrachtrup[1]

[1] *3. Physikalisches Institut, Universität Stuttgart, 70550 Stuttgart, Germany*

[2] *Department of Electrical and Computer Engineering, Texas A&M University, College Station, USA*

**Abstract.** Photon interference among distant quantum emitters is a promising method to generate large scale quantum networks. Interference is best achieved when photons show long coherence times. For the nitrogen-vacancy defect center in diamond we measure the coherence times of photons via optically induced Rabi oscillations. Experiments reveal a close to Fourier transform (i.e. lifetime) limited width of photons emitted even when averaged over minutes. The projected contrast of two-photon interference (0.8) is high enough to envisage the applications in quantum information processing. We report 12 and 7.8 ns excited state lifetime depending on the spin state of the defect.

Coherent control of single quantum systems and generation of non-classical states has attracted widespread attention because of application in quantum physics and quantum information science. Solid state systems are often considered to be promising and also difficult because of inhomogeneities and fast dephasing. Spins in solids, for example associated with quantum dots or single dopant atoms offer promising figures of merit for both parameters. As a particular example, the Nitrogen-Vacancy (NV) defect in ultrapure diamond shows long spin phase memory time (0.35 ms) [1] even at ambient conditions due to spin-free and rigid lattice. In addition to its excellent spin properties, spin selectivity of optical transitions of the NV defect allows initialization and readout of spin state with sensitivity routinely reaching single atom [2].

The narrow spin resonance transitions of single defects make them a sensitive magnetometer at the nanoscale. As an example it was demonstrated that the electron spin associated with a single NV defect can be used for reading out spin states of proximal nuclear [3] and electron spins [1,4,5]. Magnetic coupling between electron spin of NV defects and neighboring nuclei was used as a resource for generating entangled states[6]. Such multi-spin entanglement is a crucial element for quantum computation and communications protocols [7]. However, the generation of entanglement using magnetic dipolar coupling is limited to closely spaced spins. The maximum distance where spin-spin interaction can be used for controlling non-local quantum states depends on coherence time and strength of spin-spin interaction. Although coherences associated with electron and nuclear spins in diamond are particularly long, a few nanometers distance is a realistic limit for magnetic coupling. One way to gain entanglement over larger distance is to use the coupling of spin state to optical transitions [8,9]. Such generation of entanglement over long distance via interference of photons recently attracted considerable interest [10]. Experimental demonstration of entanglement between stationary and flying qubits [11,12] followed by realization of interference of photon pairs from distant trapped ions [13] and their entanglement via photonic channel [14] is an important breakthrough for modern quantum physics.

The generation of two-photon interference in solid state systems remains challenging. The major difficulties that solid state systems facing are the coherence properties of photons and the inhomogeneous distribution of transition frequencies. By using appropriate control parameters, the inhomogeneity of a solid state system can be eliminated. For single NV defects tuning of transition frequencies has been reported previously [15]. However dynamic inhomogeneity (variation of optical transition frequency related to different relaxation processes in solids) leads to mapping of environmental fluctuations into the frequency of photons. This makes two photons emitted by nominally identical systems at least in principle distinguishable destroying the contrast of the two-photon interference. Despite those challenges, several well-isolated solid state systems have been investigated in details [16,17]. In the above mentioned context we address a single NV defect in diamond. NV defect was proven to be an efficient source of broadband photons at ambient conditions [18,19] allowing spectacular experiments of single photon interference [20]. Here we present experimental evidence of nearly transform-limited transition of single NV centers at cryogenic temperature. This is proven in two steps, first by measuring the excited state lifetime for different spin states and subsequently generating optical Rabi oscillations. Surprisingly our experiment reveals new insights into the photophysics of the NV defects allowing to achieve high visibility of ESR Rabi oscillation at ambient conditions.

The structure and energy levels of the NV defect in diamond are shown in Figure 1. The defect consists of a substitutional nitrogen atom and a vacancy in nearest neighbor lattice position. The negatively charged NV center, which is relevant for our experiments, comprises six electrons two of which are unpaired. The ground $^3$A state is a spin triplet with zero-field splitting of 2.88 GHz between $m_S$=0 and $m_S$=±1 states. The structure of the excited $^3$E state is governed by spin–orbit and spin–spin interactions [21]. It is well established that at least one singlet state ($^1$A) is lying between the ground and excited triplet state [22]. Transitions between triplet and singlet states govern the spin polarization dynamics of the NV defect. It was demonstrated experimentally that the intersystem crossing (ISC) transitions $^3$E→$^1$A are strongly spin selective with the shelving rate from the $m_S$=0 sublevel being much smaller than those from $m_S$=±1 sublevels. The rates of ISC transitions towards the ground state triplet $^1$A→$^3$A are also spin selective, but here the $m_S$ =0 state is mostly populated. Hence after a few optical excitation emission cycles a strong spin polarization of the ground-state $m_S$=0 spin sublevel is established.

As a first step the excited state lifetime of the NV center was measured. Efficient optical pumping leads to spin polarization into the $m_S$=0 sublevel as discussed above. Hence it is difficult to measure the decay of the excited state $m_S$=±1 sublevels directly. It was suggested to use very low repetition rate lasers to allow the system to reach thermal equilibrium between optical pulses [22]. The long reported $T_1$ time of the spin state [23] (265 s) makes such an experiment unfeasible for single atoms. In order to detect the decay of the excited state we have chosen a different experimen-tal approach described in Figure 2. The NV center was first polarized into the $m_S=0$ state by a microsecond optical pulse. After that the fluorescence decay was measured using sub-picoseconds excitation pulses which were synchronized with single photon counting data acquisition. For measuring the decay of the $m_S=\pm1$ states a microwave π pulse was introduced after the spin polarization pulse.

Results of such measurements are shown in Figure 2. Fluorescence decay curves follow single exponential decay with time constants of 12.0 ns for $m_S=0$ state and 7.8 ns for $m_S=\pm1$ respectively. The decay of the $m_S=0$ state found here is in good agreement with a previously reported lifetime [24] whereas its dependence on the spin quantum number is surprising at the first sight. It is remarkable that the total number of photons emitted from excited $m_S=\pm1$ states is smaller compared to $m_S=0$ state. This indicates that the difference in lifetime is due to the $m_S$ dependent non-radiative passage into the metastable state and not to the change of the transition oscillator strength. The total excited state decay rate is the sum of the radiative decay rate and intersystem crossing rate, $k_{fl.}=k_{3E-3A}+k_{3E-1A}$. Since $k_{3E-1A}$ is different for $m_s=0$ and $\pm1$, $k_{fl}$ depends on the spin state. From our experiments a difference in the inter system crossing rate $k_{3E-1A}$ among the spin states of $4.5\cdot10^{-7}$ $s^{-1}$ can be deduced which is in excellent agreement with recent theoretical predictions [22].

Mentioned above finding allows for an improvement of single spin experiments. The decay into metastable state defines the contrast in cw and pulsed optically detected ESR experiments. Under stationary conditions this contrast never exceeds 30 percent. The significant difference in fluorescence lifetimes can be used for improving the optical readout of the spin state. This can be made by gating the detection channel and selecting only the long lifetime photons. Figure 3 shows the result of such selection using a gating window of 10 ns with a 100 ns delay after the laser pulse. The visibility is improved by a factor of two when compared to the previously reported scheme [2,25]. The visibility of the ESR Rabi oscillations can be improved to an arbitrary value by making delay time longer. The price to pay is loss of the count rate leading to a longer integrating time.

At low temperatures not only the lifetime, but also the frequency of photons associated with specific transition carry information about the spin state [26]. This spin-photon frequency correspondence would potentially allow entangling of two distant NV defects using photon interference. In order to demonstrate that such remote entanglement via projective state measurement is applicable to NV diamond, the coherence time of photons should be as long as possible. It was shown that nitrogen-free samples usually show narrow spectral lines. Recently lifetime-limited excitation lines were recorded under weak laser illumination [15]. However the question about the possibility to observe Fourier-transform limited photons under strong optical excitation remains to be addressed.

The figure of merit for photons in the interference experiments mentioned above is the Fourier transform relation between spectral $\Delta \nu$ and temporal $\Delta \tau$ profiles $\Delta \nu \, \Delta \tau = \frac{1}{2\pi}$. Both parameters are related to the coherence $T_2$ and relaxation $T_1$ time of the corresponding optical transitions via $\Delta \nu = \frac{1}{2\pi T_2}$ and $\Delta \tau = 2T_1$. Hence fully coherent wave packets are satisfying the criterion $\frac{2T_1}{T_2} = 1$. Several solid state systems are approaching this limit with achieved ratios of $\frac{2T_1}{T_2} = 1.6$ for single organic molecules [17] and $\frac{2T_1}{T_2} = 1.5$ for semiconductor quantum dots [27].

In order to characterize the coherence properties of the optical transitions we have used fluorescence excitation spectroscopy. A narrow single frequency tunable dye laser (linewidth 500 KHz) is tuned across the electronic transition of NV and the intensity of the broad Stokes-shifted emission into the phonon sideband is recorded. Although these red shifted wide in frequency photons cannot be used in quantum computation experiments directly, they carry important information about the coherence properties of transition. In general, the homogeneous linewidth can be defined as $\Gamma_2 = \frac{1}{T_2} = \frac{1}{2T_1} + \frac{1}{T_2^*}$. Here $T_2^*$ is the pure dephasing time. Such dephasing can be considered as source of losses in two-photon interference experiments. If dephasing is present, the depth of the Hong-Ou-Mandel dip is reduced to $\frac{T_2}{2T_1}$ [28]. Thus, it is of crucial importance to characterize $T_2^*$ of the photon source. Since under realistic conditions a single photon source has to be driven strongly, $T_2^*$ must be measured under saturating optical excitation. A single transition associated with the long-cycling $m_S$ =0 state is visible in the fluorescence excitation spectrum of the defect [21]. The derivation of $T_2^*$ from the linewidth of the detected transition is complicated due to the saturation broadening, whereas the second-order intensity correlation function of the emitted photons $g^{(2)}(\tau) = \langle I(t)I(t+\tau)\rangle / \langle I(t)\rangle^2$ allows extracting it in a more straightforward way. The experimental measurements of $g^{(2)}(\tau)$ (Figure 4) show the signature of optically induced Rabi oscillations with a decay defined by the radiative lifetime of the excited state and pure dephasing. Since the metastable state population occurs on a long timescale for the $m_S$ =0 state, the system can be analyzed in terms of optical Bloch equations for a two-level system. The corresponding second-order intensity correlation function can be written as [29]

$$g^{(2)}(\tau) = 1 - \exp\left(-\left(\frac{3}{4T_1} + \frac{1}{2T_2^*}\right)\cdot\tau\right)\cdot\left(\left(\frac{3}{4T_1} + \frac{1}{2T_2^*}\right)\frac{1}{\Omega}\sin(\Omega\tau) + \cos(\Omega\tau)\right). \tag{1}$$

Here $\Omega$ is the optical Rabi frequency. Fit functions with $T_2^*$ as fitting parameter ($T_1$ is a fixed parameter measured independently as described above) are presented together with experimental data. There is a clear indication that dephasing occurs at hard driving laser field. Note that at 75 MHz driving frequency the pure dephasing time exceed 80 ns (lower curve), which correspond to 77 % of Mandel dip depth for a two-photon interference experiment. It is important to mention that the measured $T_2^*$ is characteristic for the whole acquisition time of the experiment (a few tens of minutes). During this time approximately $10^{10}$ photons were emitted and $3\cdot10^7$ were detected. Previously reported two photon interference experiments on single quantum dots rely on indistinguishability of two consecutive photons separated by a time interval of a few nanoseconds [27]. Slow spectral jumps would probably destroy coalescence contrast on longer timescale, which is not the case for NV centers.

In summary, we show an important new insight into the coherence properties of NV centers in diamond. The NV defects show narrow linewidth of optical transition with minor contribution from pure dephasing when compared to broadening via radiative decay. To obtain single photon stream for two-photon interference experiments, the NV center must be excited non-resonantly and the photons associated with the zero-phonon line must be selected for detection. Alternatively, fast optical switches can be used to separate the excitation pulse from the fluorescence photons when resonant excitation is used. The novel contrast mechanism for single spin ESR at ambient condition based on a temporal filtering of photons might help to facilitate the detection and spin manipulations in the quantum register.

**Acknowledgement**

This work supported in part by DFG (projects "SFB/TRR 21" and "WR 26/16"), EU (QAP, EQUIND, NEDQIT) and foundation "Landesstiftung B-W" (project "Atomoptik").

**Figure captions**

FIG. 1 (color online). (a) Structure of the nitrogen-vacancy (NV) center in diamond. The NV center comprises a substitutional nitrogen (N), and a neighboring vacancy (V). (b) The scheme of energy levels of the NV defect center. Thick arrows indicate spin-selective excitation and fluorescence emission pathways (transitions a,b and c).

FIG. 2 (color online). Spin-selective decay curves of a single NV defect at ambient conditions. The inset shows the time sequence of the experiment. The $m_S = 0$ spin states have been initialised by a μs long optical pulse followed by long (μs) delay. Next, the center was excited by 150 fs laser pulse and the fluorescence decay was recorded (this sequence was used for lifetime measurements of $m_S=0$ state, squares, corresponds to transition $a$ shown in Figure 1). Introduction of a spin selective microwave π pulse after the polarization pulse initializes the system into $m_S=\pm1$ prior to lifetime measurements (open and filled circles, transition $b,c$ shown in Figure 1). Solid lines are single exponential fit curves.

FIG. 3 (color online). Rabi oscillations of a single NV electron spin recorded using temporal gating (solid circles). After initialization of the NV center into the $m_S = 0$ state a resonant microwave pulse of variable length was applied. Consecutively a short (100 fs) pulse was used for reading out. In order to make the detection scheme sensitive to the $m_S = 0$ state population, only photons with long lifetime were selected by choosing the delay of detection window to be 100 ns. Measurement based on discrimination of spin states via total fluorescence is shown for comparison (open circles).

FIG. 4 (color online). Second order fluorescence intensity autocorrelation function for the NV center at low temperature under resonant excitation of $m_S = 0$ spin state (corresponds to transition *a* shown in Figure 1). The Rabi frequency of the excitation laser was varied for different curves (increase from bottom to top). The fit functions (solid lines) are based on eq. 1 (see text).

*Electronic address: f.jelezko@physik.uni-stuttgart.de

**References**

[1] T. Gaebel *et al.*, Nature Physics **2**, 408 (2006).
[2] F. Jelezko *et al.*, Physical Review Letters **92**, 076401 (2004).
[3] L. Childress *et al.*, Science **314**, 281 (2006).
[4] R. J. Epstein *et al.*, Nature Physics **1**, 94 (2005).
[5] R. Hanson *et al.*, Physical Review Letters **97** (2006).
[6] M. V. G. Dutt *et al.*, Science **316**, 1312 (2007).
[7] P. Zoller *et al.*, European Physical Journal D **36**, 203 (2005).
[8] L. Childress *et al.*, Physical Review A **72**, 052330 (2005).
[9] L. Childress *et al.*, Physical Review Letters **96**, 070504 (2006).
[10] S. D. Barrett, and P. Kok, Physical Review A **71**, 060310 (2005).
[11] B. B. Blinov *et al.*, Nature **428**, 153 (2004).
[12] J. Volz *et al.*, Physical Review Letters **96** (2006).
[13] P. Maunz *et al.*, Nature Physics **3**, 538 (2007).
[14] D. L. Moehring *et al.*, Nature **449**, 68 (2007).
[15] P. Tamarat *et al.*, Physical Review Letters **97**, 083002 (2006).
[16] C. Santori *et al.*, Physical Review Letters **97**, 247401 (2006).
[17] A. Kiraz *et al.*, Physical Review Letters **94** (2005).
[18] R. Brouri *et al.*, Optics Letters **25**, 1294 (2000).
[19] C. Kurtsiefer *et al.*, Physical Review Letters **85**, 290 (2000).
[20] V. Jacques *et al.*, Science **315**, 966 (2007).
[21] P. Tamarat *et al.*, arXiv **cond-mat/0610357** (2006).
[22] N. B. Manson, and R. L. McMurtrie, Journal of Luminescence **127**, 98 (2007).
[23] J. Harrison, M. J. Sellars, and N. B. Manson, Diamond and Related Materials **15**, 586 (2006).
[24] A. T. Collins, M. F. Thomaz, and M. I. B. Jorge, Journal of Physics C-Solid State Physics **16**, 2177 (1983).
[25] F. Jelezko *et al.*, Physical Review Letters **93**, 130501 (2004).
[26] F. Jelezko *et al.*, Applied Physics Letters **81**, 2160 (2002).

[27]   C. Santori *et al.*, Nature **419**, 594 (2002).
[28]   J. Bylander, I. Robert-Philip, and I. Abram, European Physical Journal D **22**, 295 (2003).
[29]   T. Basche *et al.*, Physical Review Letters **69**, 1516 (1992).

Batalov et al., Figure 1

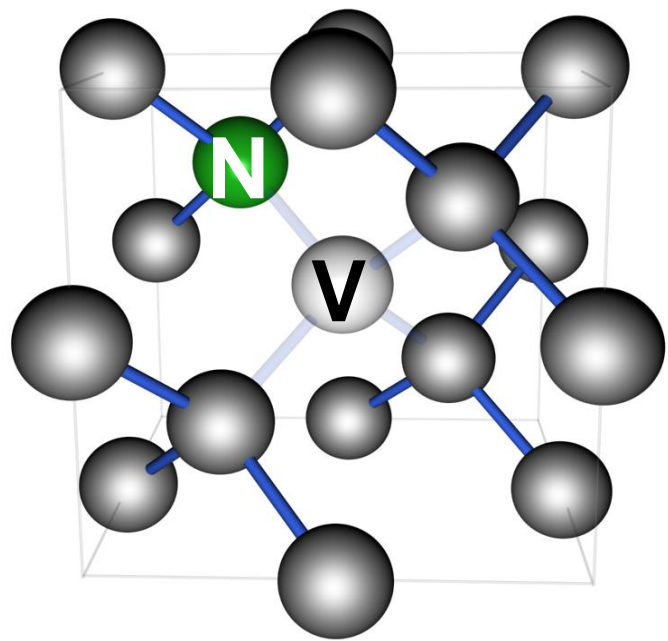
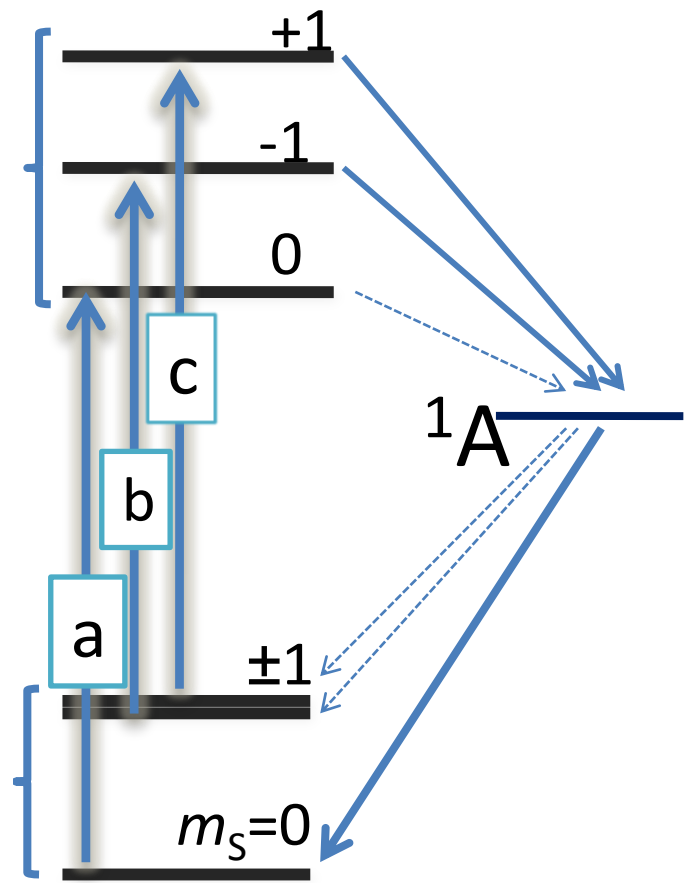

(a) (b)

Batalov et al., Figure 2

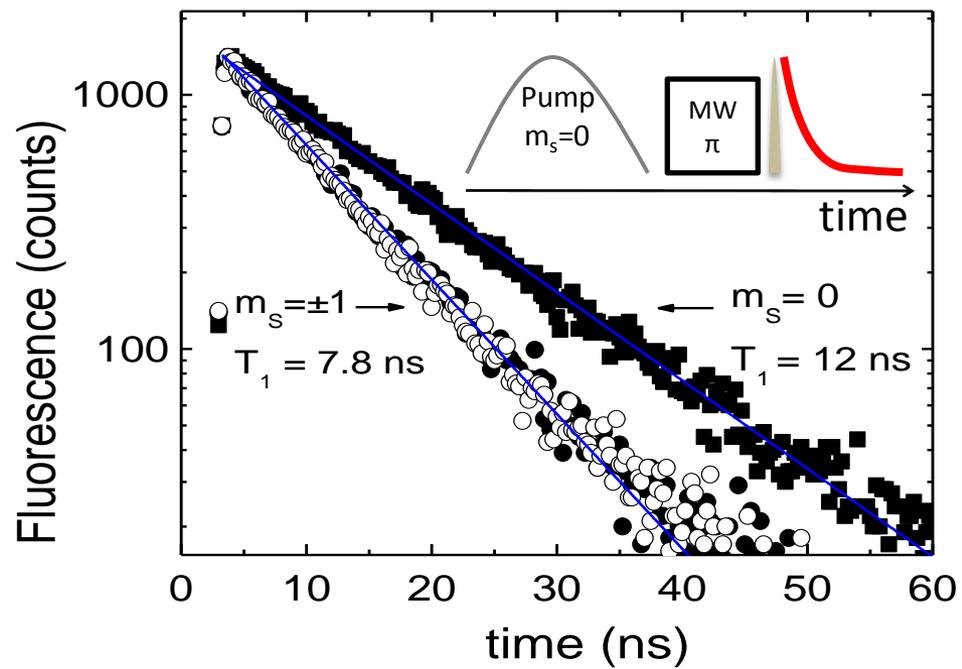

Batalov et al., Figure 3

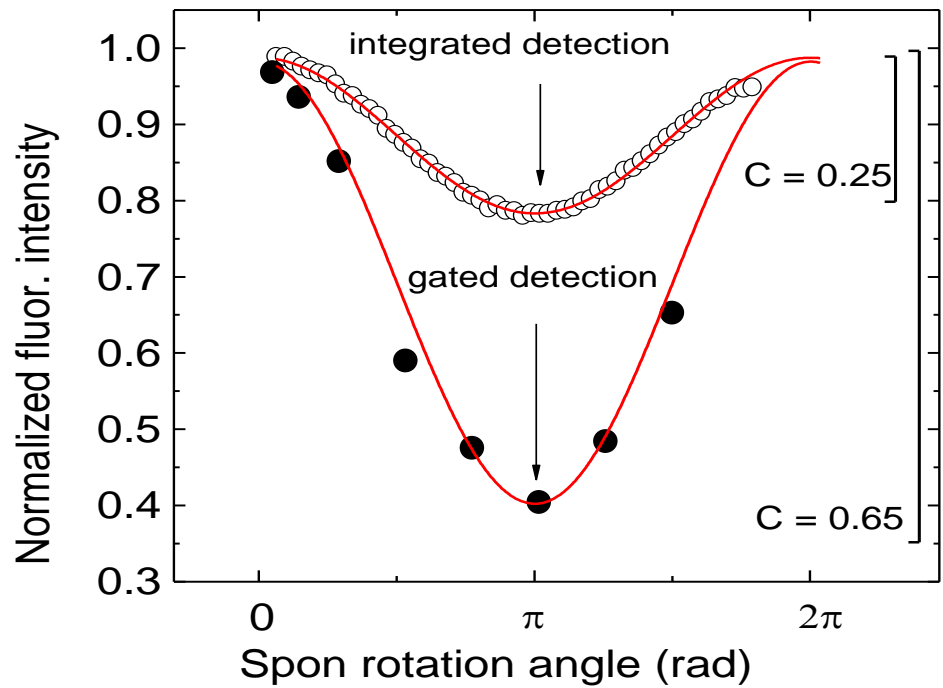

Batalov et al., Figure 4

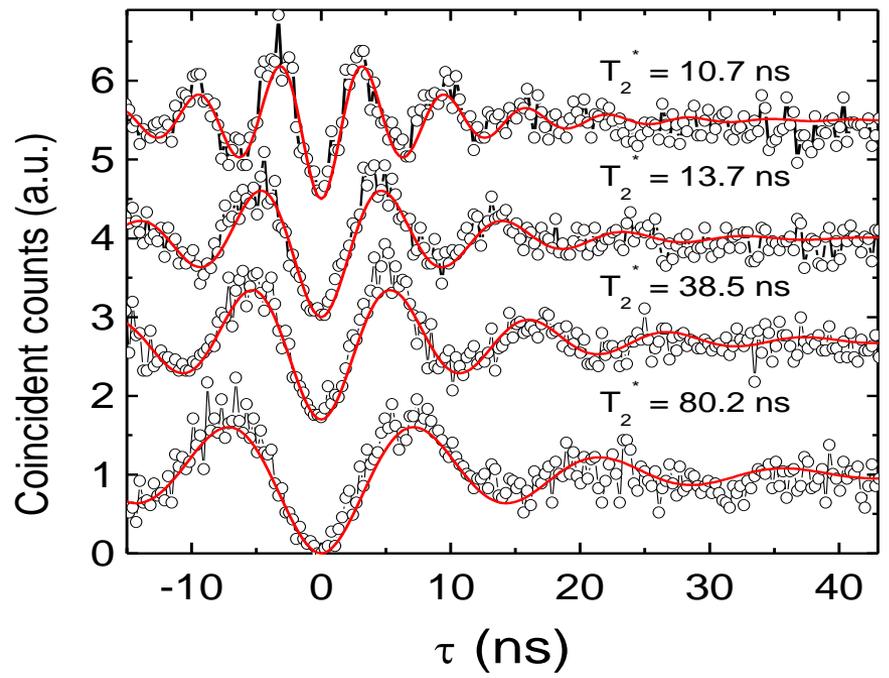